\documentclass[11pt,titlepage]{article}

\def\N{{\rm I\!N}} 
\newcommand{\I}{{\displaystyle{\bf i}}}
\newcommand{\G}{\mbox{$\cal G\ $}}
\newcommand{\A}{\mbox{$\cal A\ $}}
\newcommand{\B}{\mbox{$\cal B\ $}}
\newcommand{\F}{\mbox{$\cal F\ $}}

\newtheorem{mytheorem}{Theorem}

\begin{document}
\title{UNREAL PROBABILITIES \\
\emph{Partial Truth with Clifford Numbers}}
\author{Carlos C. Rodriguez\\
Department of Mathematics and Statistics\\
University at Albany, SUNY\\
Albany NY 12222, USA\thanks{carlos@math.albany.edu}}
\maketitle
\tableofcontents
\begin{abstract}
This paper introduces and studies the basic properties of Clifford algebra
valued conditional measures.

\end{abstract}

\section{Introduction}

Probability theory was given a firm mathematical foundation in 1933,
when Kolmogorov \cite{kolmogorov33} introduced his axioms. By defining
probability as an \emph{uninterpreted} special case of a positive
measure with total unit mass (plus an additional definition for
independence), the subject exploded with new results and
found innumerable applications. In 1946, Cox (see \cite{cox46}) showed
that the Kolmogorov axioms for probability are really theorems that
follow from basic desiderata about the representation of partial truth
with real numbers. We owe to Ed Jaynes (see \cite{jaynesBook}) the
discovery of the importance of Cox's 1946 work (\cite{cox46}). After
Jaynes, it became clear why the calculus of probability is so
successful in the real world. Probability works because its axioms
axiomatize the right thing: \emph{partial truth of a logical
proposition given another}. Even more, the rules of probability are
unique in the sense that any other set of consistent rules can be
brought into the standard sum and product rules by a change of scale
(or we may say logical gauge). This is in fact Cox's main result and
it makes futile the enterprise of looking for alternatives to the
calculus of normalized real valued probabilities.  It is only by
allowing the partial truth of a proposition to be encoded by an object
other than a real number in the interval $[0,1]$ that we could find
alternatives to the standard theory of probability.

We seek to find out what happens when standard probability theory is
modified by relaxing the axiom that the probability of an event must
be a real number in the interval $[0,1]$.  We show that, by allowing
the measure of a proposition to take a value in a Clifford Algebra, we
automatically find the methods of standard quantum theory without ever
introducing anything specifically related to nature itself.

The main motivation for this article has come from realizing that the
derivations in Cox \cite{cox46} still apply if real numbers are
replaced by complex numbers as the encoders of partial truth. This was
first mentioned by Youssef \cite{youssef91} and checked in more detail
by Caticha \cite{caticha98} who also showed that non-relativistic
Quantum theory, as formulated by Feynman \cite{feynman48}, is the only
consistent calculus of probability amplitudes.  By measuring
propositions with Clifford numbers we automatically include the reals,
complex, quaternions, spinors and any combination of them (among
others) as special cases.

\section{The Axioms}

In this section we introduce the notation and collect the simple properties
about Boolean and Clifford algebras that will be needed for the definition
of $\psi$ below.

\subsection{The Boolean Algebra \mbox{$\cal A\ $}}

Let \A be a boolean $\sigma$-algebra of propositions $%
a,b,c,\ldots$. We denote by 0 the false proposition, by 1 the true
proposition, by $a+b$ the logical sum, by $ab$ the logical product and by $%
\bar{a}$ the negation. Each proposition $b\in\A$ defines the
set $\A_{b}$ where, 
\begin{equation}
\A_{b} = \{ ba : a\in\A \} = b\A
\label{eq:Ab1}
\end{equation}
Clearly, $\A_{b}$ is a subset of \A that
contains $b$ and $0$, it is closed for sums and products and thus, $A_{b}$
is a sub algebra of \A with $b$ as the unit. From the fact
that $a = ac + a\bar{c}$ it follows that 
\begin{equation}
\A = \A_{c} \oplus \A_{\bar{c}}
\label{eq:AplusA}
\end{equation}
Given two propositions $a,b\in\A$ we have, 
\begin{equation}
\A \supset \A_{a} \supset \A%
_{ab}  \label{eq:Anested}
\end{equation}
The set $X\subset\A$ is called the set of elementary
propositions of \A (and we say that \A is a $%
\sigma$-algebra of propositions in $X$) if,

\begin{enumerate}
\item  for $x,y\in X$, $xy=0$ whenever $x\neq y$

\item  every $a\in \A$ is the sum of propositions in $X$. We
write 
\begin{equation}
a=\sum_{x\in a}x  \label{eq:asumx}
\end{equation}
\end{enumerate}

If \A and $\mathcal{B}$ are two Boolean $\sigma$-algebras of
propositions in $X$ and $Y$ respectively, then $\A\times 
\mathcal{B}$ is a Boolean $\sigma$-algebra of propositions in $X\times Y$ if
one defines the truth value of $(a,b)\in\A\times \mathcal{B}$
as the truth value of $ab$ i.e. true only when both $a\in\A$
and $b\in\mathcal{B}$ are true. We denote by $\A^{n}$ the $%
\sigma$-algebra of $n$ copies of \A of propositions in $X^{n}$%
. We have, 
\begin{equation}
P\in \A^{n} \Longleftrightarrow P = \sum_{x\in
P}x_{1}x_{2}\ldots x_{n}  \label{eq:An}
\end{equation}
and by this we mean that $P$ is always the sum of propositions in $X^{n}$
and each $x\in X^{n}$ is the conjunction of $n$ propositions, one for each
copy of $X$. Finally we let $\A^{*} = \A
\setminus \{0\}$. 

Notice that these Boolean $\sigma$-algebras are nothing but the standard
sets where general measures (in particular probability measures) are
defined. We chose the notation of logical sums and products instead 
of the traditional set notation of unions and intersections to emphasize
the fact that we are interested in the encoding of partial truth of 
logical propositions, but this is only a choice of notation and 
there is a complete one to one correspondence between the two languages.
As general references see e.g. Halmos \cite{halmos74} or
Chow and Teicher \cite{chow88}.

\subsection{The algebra of Clifford numbers \G}

Let \G be an arbitrary finite dimensional Clifford Algebra
with real scalars. We try to follow the notation in \cite{hestenes84}. We
denote the elements of \G by capital letters like, $%
A,B,C,\ldots$. A general Clifford number $M$ always expands as the sum of
its scalar, vector, bivector, etc parts like: 
\begin{eqnarray}
M & = & <M>_{0} + <M>_{1} + <M>_{2} + \ldots  \label{eq:M} \\
& = & \alpha + u + B + \ldots  \nonumber
\end{eqnarray}
Where $<M>_{k}$ denotes the k-vector part of $M$. If $u$ and $v$ are vectors
in \G then their geometric (Clifford) product $uv$ can be
decomposed into a symmetric part $u\cdot v$ and an antisymmetric part $%
u\wedge v$ as 
\begin{eqnarray}
uv & = & \frac{1}{2}(uv + vu) + \frac{1}{2}(uv - vu)  \label{eq:uv1} \\
& = & u\cdot v + u\wedge v  \label{eq:uv2}
\end{eqnarray}
The inner product between two vectors is always a scalar and their wedge
product is always a bivector. The operation of \emph{reversion} of a
clifford number $M$ is denoted by $M^{\dagger}$ and defined as a linear
operation with the properties, 
\begin{equation}
\alpha^{\dagger} = \alpha,\ \ u^{\dagger} = u,\ \ (MN)^{\dagger} =
M^{\dagger}N^{\dagger}  \label{eq:dagger}
\end{equation}
where $\alpha$ is a scalar, $u$ is a vector, and $M$ and $N$ are arbitrary
Clifford numbers. The euclidean inner product on \G is given
by, 
\begin{equation}
<M,N>_{\G} = <M^{\dagger}N>_{0}  \label{eq:inner}
\end{equation}

\subsection{Definition of $\psi$}

By a clifford algebra valued conditional measure (or simply a $\psi$) we
mean a function, 
\begin{equation}
\begin{array}{cccc}
\psi : & \A \times \A^{*} & \longmapsto & %
\G \\ 
& (a,c) & \longmapsto & \psi(a,c)
\end{array}
\end{equation}
satisfying the following two axioms:

\begin{description}
\item[(I)]  If $c \Rightarrow b$ then 
\begin{equation}
\psi(a,c) = \psi(ab,c)  \label{eq:psiI}
\end{equation}

\item[(II)]  If $\{a_{1},a_{2},\ldots\} \subset \A$, with $%
a_{j}a_{k} = 0$ for $j \neq k$, then 
\begin{equation}
\psi\left(\sum_{j}a_{j},c\right) = \sum_{j}\psi(a_{j},c)  \label{eq:psiII}
\end{equation}
\end{description}

Since the only property a proposition in \A always has is its truth
value, we can interpret $\psi(a,c)$ as the clifford number that
represents the truth in $a$ when $c$ is certain. Axiom (I) says that
$c$ is certain (e.g. take $a=1$ and $b=c$) and axiom (II) says that the
whole truth of $a$ for a given $c$ is always the sum of the truths of
its separate parts.

\subsubsection{The truth of $0$}

By taking each $a_{j}=0$ in (\ref{eq:psiII}) we get, 
\begin{equation}
\psi(0,c) = \psi(0,c) + \psi(0,c) + \ldots
\end{equation}
and therefore, $\psi(0,c)$ is either $0\in \G$ or unbounded
but if it is unbounded then all the propositions will be assigned an
unbounded value since $\psi(a,c) = \psi(a+0,c) = \psi(a,c) + \psi(0,c)$.
Hence, 
\begin{equation}
\psi(0,c) = 0 \ \ \mbox{for all } c \in \A^{*}
\label{eq:psi0}
\end{equation}

\section{The spaces $H_{c}$}

The functions $\psi$, as defined by (\ref{eq:psiI}) and (\ref{eq:psiII}),
are specified independently at each $c\in\A^{*}$. So far,
there is no link between the $\psi$ in the domain of discourse of $c$, i.e. $%
\psi(\cdot,c)$ and $\psi$ in the more specialized domain of discourse of $bc$%
, i.e. $\psi(\cdot,bc)$. We shall talk about changing domains of discourse
in the next section but in this section we describe the important properties
that the functions $\psi(\cdot,c)$ have as functions of their first argument
only, for fixed $c\in\A^{*}$. To simplify the notation simply
write $\psi(a)$ instead of $\psi(a,c)$ in the formulas below. Thus, whenever
the background proposition $c$ is not subject to change we take $\psi$ as
any $\sigma$-additive function defined on $\A_{c}$ with
values in $\G$. The condition (\ref{eq:psiI}) is
automatically satisfied since $c$ is the true proposition in $\mbox{$\cal A\
$}_{c}$.

Let $H_{c}$ be the set of all $\sigma$-additive functions defined on
$\A_{c}$ with values in \G.

\subsection{The $H_{c}$ are Hilbert spaces}

Since the sum of two $\sigma$-additive functions and the product of a
$\sigma $-additive function by a scalar are still $\sigma$-additive
functions we have that the $H_{c}$ are vector spaces. The scalars are
the scalars in \G. In principle the field of scalars could be taken as
the reals or the complex numbers but it seems that the reals is all
that is needed in most applications.

\subsubsection{The inner product in $H_{c}$}

For, $\varphi,\psi\in H_{c}$ define the real inner product between them by: 
\begin{eqnarray}
<\varphi,\psi> & = & \sum_{x\in X} <\varphi(x),\psi(x)>_{\G}
\\
& = & \sum_{x\in X} <\varphi(x)^{\dagger}\psi(x)>_{0}  \label{eq:innerprod}
\end{eqnarray}
By considering only $\psi$s with finite norm we make $H_{c}$ a real Hilbert
space. From now on we assume the finite norm to be part of the definition of 
$H_{c}$ itself, i.e. 
\begin{equation}
H_{c} = \{\psi : \psi \mbox{\ is\ } \sigma-\mbox{additive on\ } \mbox{$\cal
A\ $}_{c} \mbox{\ and\ } \sum_{x\in X} <\psi^{\dagger}(x)\psi(x)>_{0} %
\mbox{\ } < \infty \}  \label{eq:Hc1}
\end{equation}

Notice that the spaces $H_{c}$ are complete for the inner product
(\ref{eq:innerprod}) since $\G$ with the scalar product $<.,.>_{\G}$
is complete. When $X$ is a finite set (i.e. when it contains only a
finite number of propositions) the proof is trivial, just use the fact
that if $\{\phi_{n}\}_{n\in\N}\subset H_{c}$ is a Cauchy sequence then for
each $x\in X$ the sequence $\{\phi_{n}(x)\}_{n\in\N}$ is also a Cauchy
sequence of elements of $\G$ and thus it converges to some
$\phi(x)\in\G$ and therefore $\phi\in H_{c}$ is the limit of the
original sequence in $H_{c}$.  When $X$ is infinite we need to
reinterpret the sums as integrals, (for which we need a measure in
$X$), and we also need to reinterpret the $\psi$s as $\A$-measurable
densities, but after that, the proof is essentially the standard proof
that $L^{2}$ is complete. An important example of an infinite $X$
occurs when the propositions in $X$ are labeled with the vectors in
$\G$. In this case the sum in (\ref{eq:innerprod}) is replaced by the
integral with respect to the standard Lebesgue measure in $X$.

\subsection{The isomorphic spaces: $H(\A_{c}) \simeq H_{c}(
\A)$}

In order to be able to understand the differences between the current
approach and ordinary probability theory it is convenient to introduce two
other spaces closely related to $H_{c}$. These are, the space of all $\sigma$%
-additive functions on $\A_{c}$ with values on \mbox{$\cal G\
$},

\begin{equation}
H(\A_{c}) = \{\psi_{c}:\A_{c} \longrightarrow %
\G | \psi_{c} \mbox{\ $\sigma$-additive on\ } \mbox{$\cal A\
$}_{c} \}  \label{eq:HAc}
\end{equation}
and the space, 
\begin{equation}
H_{c}(\A) = \{\psi:\A \longrightarrow %
\G | \psi \mbox{\ $\sigma$-additive on\ } \A %
\mbox{\ AND if\ } c \Rightarrow b, \psi(ab) = \psi(a) \}  \label{eq:HcA}
\end{equation}
Both are Hilbert with the inner product (\ref{eq:innerprod}) and considering
only elements of finite norm.

Notice that if $\psi\in H_{c}(\A)$ then its restriction to $%
\A_{c}$ belongs to $H(\A_{c})$, i.e. 
\[
\psi|_{\A_{c}} = \psi_{c} \in H(\A_{c}) 
\]
and conversely, if $\psi_{c}\in H(\A_{c})$ then the function $%
\psi$ defined by: 
\[
\psi(a) = \psi_{c}(ac) \ \forall a\in\A 
\]
belongs to $H_{c}(\A)$ since it is clearly $\sigma$-additive
and if $c\Rightarrow b$ then $\bar{c}+b=1$ and multiplying both sides by $c$
we get, $bc=c$ from where, 
\[
\psi(ab) = \psi_{c}(abc) = \psi_{c}(ac) = \psi(a) 
\]
The map $\psi \rightarrow \psi_{c}$ is obviously linear one to one and onto
so it makes the two spaces isomorphic.

Consider now two propositions $b$ and $c$ such that $c, bc \in\mbox{$\cal A\
$}^{*}$. Then we can write: 
\begin{equation}
H(\A_{bc}) \simeq H(b\A_{c}) \simeq H_{b}(%
\A_{c})  \label{eq:HHH}
\end{equation}
In other words each $\psi(\cdot,bc)\in H(\A_{bc})$ uniquely
defines a function $\varphi(\cdot b,c)\in H_{b}(\A_{c})$ and
that is all we can say. Since the $\psi$s are unnormalized we can not write
a general product rule as in normalized standard probability theory.
Nevertheless, it is possible to justify a restricted product rule for
independence as we do in section 5 below. When $c=1\in \A$ we simply
write $H(\A)$ instead of $H_{1}(\A)$ or $H(\A_{1})$.

\section{The truth with $\psi$}

The remarkable fact about the functions $\psi$ is that without
committing to a particular value for $\psi(1)$ in \G, they still allow
to tell what propositions are true. We show in this section that,
$b\in \A$ is considered to be true by $\psi$ when $\psi(\bar{b}a) = 0\ 
\forall a\in\A $. By liberating ordinary probabilities from the
constrain that the probability of the whole space must always be fixed
at one, we make the space of all possible assignments of partial truth
into a Hilbert space without losing the ability to identify truth.

\subsection{Propositions as operators}

Each proposition $b\in\A$ defines two complementary linear
operators on $H_{c}$ by multiplication, $\hat{b}$, and by addition, $\check{b%
}$ to the first argument of $\psi$. In symbols 
\begin{eqnarray}
\hat{b}\psi(a,c) & = & \psi(ab,c) \ \forall a\in\A
\label{eq:bhat} \\
\check{b}\psi(a,c) & = & \psi(a+b,c)\ \forall a\in\A
\label{eq:bchek}
\end{eqnarray}
To simplify the notation we often omit the hats and simply write
$b\psi$ instead of $\hat{b}\psi$.  From $bb=b$ and $b+b=b$ it follows
that $\hat{b}$ and $\check{b}$ are projectors and therefore they are
self-adjoint with eigen values either $0$ or $1$. We can write,

\begin{mytheorem}
The following two complementary statements are true.

\begin{enumerate}
\item If $\psi\in H_{c}$ is an eigen vector of the operator $\hat{b}$
with eigen value $1$ then $\psi(b) = \psi(1)$ and we say that $\psi$
makes $b$ true conditional on $c$.  Conversely, if $c\Rightarrow b$
then every $\psi\in H_{c}$ is an eigen vector of the operator
$\hat{b}$ with eigen value $1$.

\item If $\psi\in H_{c}$ is an eigen vector of the operator
$\check{b}$ with eigen value $1$ then $\psi(b) = 0$ and we say that
$\psi$ makes $b$ false conditional on $c$ . Conversely, if
$c\Rightarrow \bar{b}$ then every $\psi\in H_{c}$ is an eigen vector
of the operator $\check{b}$ with eigen value $1$.
\end{enumerate}
\end{mytheorem}

\textbf{Proof}\newline

\begin{enumerate}
\item  
\[
\hat{b}\psi = \psi \Rightarrow \psi(ab,c) = \psi(a,c) \ \forall a\in%
\A \Rightarrow \psi(b,c) = \psi(1,c) 
\]
where the last implication follows by taking $a=1$. Conversely, 
if $c \Rightarrow b$ then from (\ref{eq:psiI}) we have, 
\[
\psi(ab,c) = \psi(a,c) \ \forall a\in\A 
\]
Thus, $\hat{b}\psi = \psi$.

\item  
\[
\check{b}\psi = \psi \Rightarrow \psi(a+b,c) = \psi(a,c) \ \forall a\in%
\A \Rightarrow \psi(b,c) = \psi(0,c) 
\]
where the last implication follows by taking $a=0$. Conversely, 
\[
(c \Rightarrow \bar{b}) \Rightarrow \bar{c} + \bar{b} = 1 \Rightarrow a+bc =
a \ \forall a\in\A 
\]
Now from the fact that $\psi$ is a function and applying (\ref{eq:psiI})
twice we have, 
\begin{eqnarray*}
\psi(a+bc,c) &=& \psi(a,c) \ \forall a\in\A  \nonumber \\
\psi(ac+bc,c) &=& \psi(a,c) \ \forall a\in\A  \nonumber \\
\psi(a+b,c) &=& \psi(a,c) \ \forall a\in\A  \nonumber \\
\end{eqnarray*}
Thus, $\check{b}\psi = \psi$ $\bullet$
\end{enumerate}
The following theorem elaborates on the same theme.

\begin{mytheorem}
Let $b\in\A$ be an arbitrary proposition in a $\sigma$-algebra of
propositions in $X$ and let $\psi\in H(\A)$. 
The following are all equivalent:
\begin{enumerate}
\item $b\psi = \psi$ i.e., $\psi$ makes $b$ true.
\item $\bar{b}\psi = 0$ i.e., $\psi$ makes $\bar{b}$ false.
\item $\|\bar{b}\psi\| = 0$
\item $\|\psi\| = \|b\psi\|$
\end{enumerate}
\end{mytheorem}
{\bf Proof:}\\
We show that $1\Leftrightarrow 2\Leftrightarrow 3\Leftrightarrow 4$.
First equivalence follows from $\psi = b\psi + \bar{b}\psi$, the second
equivalence is a property of the norm and the third equivalence is
Pythagoras theorem since $(b\psi) \perp (\bar{b}\psi)\  \bullet$

It is evident from this last theorem that the norm in the Hilbert
spaces $H_{c}$ provides a mechanism for translating the clifford
numbers $\psi(b)$ assigned to the propositions in \A by a function
$\psi\in H(\A)$ into positive real numbers 
\begin{eqnarray}
\|\bar{b}\psi\|^{2} & = & \|(1-b)\psi\|^{2} = \|\psi - b\psi\|^{2} 
	\label{eq:bbarpsi1}\\
		    & = & \|\psi\|^{2} - \|b\psi\|^{2} \label{eq:bbarpsi2}
\end{eqnarray}
measuring how close is $\psi$ from making the proposition $b$ true.
It is also clear from (\ref{eq:bbarpsi1}) and (\ref{eq:bbarpsi2})
that it is the square of the norm and not just the norm what is needed.
It is only with the square of the norms that we can say that the
amount of truth of $\bar{b}$ (measured by $\|\bar{b}\psi\|^{2}$)
equals the amount of truth assigned to the true proposition
(measured by $\|1\psi\|^{2}$) minus the amount of truth assigned to
$b$ (measured by $\|b\psi\|^{2}$).

\subsection{Commutativity, orthogonality and a Clifford number times a
proposition}

Propositional operators can be composed to form other operators. Thus, if $%
a,b\in\A$ and $\psi\in H(\A)$ we have, 
\begin{eqnarray}
(\hat{a}\hat{b})\psi(x) & = & \hat{a}(\hat{b}\psi(x)) = \psi(abx) = \hat{(ab)%
}\psi(x) \\
(\hat{a}\check{b})\psi(x) & = & \hat{a}\psi(x+b) = \psi(ax+ab) \\
(\check{b}\hat{a})\psi(x) & = & \check{b}\psi(ax)= \psi(ax+b)
\end{eqnarray}
and we can see that checks commute with other checks and hats commute with
other hats, but in general, hats don't commute with checks. 

If $A\in\G$ and $b\in\A$ we can define the
operator $Ab$ by, 
\begin{equation}
(Ab)\psi(x) = A(b\psi(x)) = A\psi(bx)  \label{eq:Ab}
\end{equation}
and similarly for $A\check{b}$. These definitions allow a very rich algebra
of operators that mix boolean and clifford algebra properties in new ways.
One particularly interesting example of this kind of mix is given by the
following statement: \emph{mutually exclusive propositions are orthogonal}.
More explicitly, if $a,b\in \A$ and $\psi_{1},\psi_{2}\in H(%
\A)$ then, $ab=0 \Rightarrow <a\psi_{1},b\psi_{2}> = 0$ and
pythagoras theorem holds, 
\begin{equation}
\| a\psi_{1} + b\psi_{2} \|^{2} = \|a\psi_{1}\|^{2} + \|b\psi_{2}\|^{2}
\label{eq:abpythagoras}
\end{equation}

\section{Independence}

If the clifford number $\psi(a,c)$ is interpreted as a representation of the
partial truth of $a$ when we assume $c$ to be certain then there is only one
rational way to define independence namely:

\textbf{Preliminary Definition:} We say that $\psi$ makes propositions $a$
and $b$ in $\A$ \emph{logically independent} conditionally on 
$c\in\A^{*}$ if the additional knowledge of one of them does
not affect the value of $\psi$ for the other. i.e, 
\begin{eqnarray}
\psi(a,bc) &=& \psi(a,c)  \nonumber \\
&AND&  \label{eq:indep} \\
\psi(b,ac) &=& \psi(b,c)  \nonumber
\end{eqnarray}
whenever the conditional $\psi$s exist.

\subsection{A restricted product rule}

If we try to find the value of $\psi(ab,c)$ in terms of the partial truths
that $\psi$ assigns to $a$ and $b$, then the most general relation is, 
\begin{equation}
\psi(ab,c) = F(\psi(a,c),\psi(a,bc),\psi(b,c),\psi(b,ac))  \label{eq:F1}
\end{equation}
where $F$ is an arbitrary function of its arguments. If we assume further
that $a$ and $b$ are logically independent conditionally on $c$ then using (%
\ref{eq:indep}) the most general relation becomes, 
\begin{equation}
\psi(ab,c) = F(\psi(a,c),\psi(b,c))  \label{eq:F}
\end{equation}
Let $u=\psi(a,c), v=\psi(b,c)$ and $w=\psi(d,c)$ and use the commutativity
and associativity properties of the logical product to get the following two
properties for the function $F$: 
\begin{eqnarray}
F(u,v) &=& F(v,u)  \label{eq:Fsym} \\
F(F(u,v),w) &=& F(u,F(v,w))  \label{eq:Fassoc}
\end{eqnarray}
In other words, $F$ must be symmetric and it must satisfy the usual
associativity equation. If the $\psi$s take values only on a
commutative subspace of \G (e.g. reals, complex or pseudo scalars)
then the only solution is $F(u,v)=uv$ (see \cite{caticha98}) but this
can not be the solution if $uv \neq vu$. Given that $F(u,v)$ must be
symmetric, and that it must reduce to $uv$ when $u$ and $v$ commute,
and obvious solution is given by the symmetrization of the product,
i.e. $(uv+vu)/2$. In principle, it seems feasible that a modification
of the standard argument of Aczel (see \cite{aczel66} or
\cite{smith90}) may yield the symmetrized product as the unique
solution of (\ref{eq:Fassoc}) and (\ref{eq:Fsym}) for $u,v\in\G$ 
at least for $u,v$ in some subset of $\G$ for which (\ref{eq:Fassoc})
is still true. At the present time there is no such proof. In any case
the lack of a proof for the uniqueness is not a deterrent to turn the
formula into the definition for independence. If it turns out that
there are multiple solutions (which seems highly unlikely) the results
obtained from this particular solution will still be valid. Thus, from
now on we say that $\psi$ makes $a$ and $b$ (logically) independent
given $c$ if,
\begin{equation}
\psi(ab,c) = \frac{1}{2}[ \psi(a,c)\psi(b,c) + \psi(b,c)\psi(a,c)]
\label{eq:symme}
\end{equation}
More generally we have, \newline
\newline
\textbf{Definition:} We say that $\psi$ makes $a_{1},a_{2},\ldots,a_{n}$
logically independent given $c$ if, for $k=1,2,\ldots,n$ and
$1\le i_{1} < i_{2} < \ldots < i_{k} \le n$
\begin{equation}
\psi(\prod_{j=1}^{k}a_{i_{j}},c) = \frac{1}{k!}\sum_{\sigma}\psi(a_{
\sigma(i_{1})},c) \psi(a_{\sigma(i_{2})},c)\ldots\psi(a_{\sigma(i_{k})},c)
\label{eq:prodrule}
\end{equation}
where the sum runs over all the permutations, $\sigma$ of 
$(i_{1},i_{2},\ldots,i_{k})$.

The associativity equation (\ref{eq:Fassoc}) imposes a heavy restriction
on the possible $\psi$ assignments for independent propositions. In fact
we have,

\begin{mytheorem}
If $\psi$ makes three or more propositions $a,b,d,\ldots$
independent conditionally on $c$ then the clifford numbers
$u=\psi(a,c), v=\psi(b,c), w=\psi(d,c),\ldots$ are such that
each of them commutes with the anticommutator of any other two.
\end{mytheorem}
{\bf Proof:}\\ 
From (\ref{eq:prodrule}) it suffices to show that
$v$ must commute with $F(u,w) = (uw+wu)/2$ when
$a,b,d$ are independent given $c$. 

The right hand side of (\ref{eq:Fassoc}) simplifies to,
\begin{equation}
F(u,F(v,w)) = \frac{1}{4}
 \left\{ uvw  - uwv + vwu - wvu \right\} \label{eq:lhsFassoc}
\end{equation}
Similarly the left hand side of (\ref{eq:Fassoc}) is given by,
\begin{equation}
F(F(u,v),w) = \frac{1}{4}
 \left\{ uvw - vuw + wuv - wvu \right\} \label{eq:rhsFassoc}
\end{equation}
Equating (\ref{eq:lhsFassoc}) to (\ref{eq:rhsFassoc}) and simplifying
we get,
\begin{equation}
[v,F(u,w)] = 0 		\label{eq:vFuw}
\end{equation}
where $[u,v] = uv - vu$ denotes the usual commutator product
$\bullet$

Notice that when the clifford numbers $u,v,w,\ldots$ either
commute or anticommute with each other then (\ref{eq:vFuw})
is true. But there are many other solutions. For example
(\ref{eq:vFuw}) is also true when $u,v,w,\ldots$
are arbitrary vectors.

\subsection{Independence and Orthogonality}

The above definition for independence makes the following statement true:

\begin{mytheorem}
If $\psi(ab,c) = 0$ then $\psi(a,c)$ anticommutes with $\psi(b,c)$ 
when and only when  $\psi$ makes $a$ and $b$ independent given $c$.
\end{mytheorem}

There is nothing like this in standard probability theory where
mutually exclusive events that are possible (i.e. that have positive
probability) are never independent. We are used to think that this
makes sense, for if we know that one of the events happens then we
also know that the other couldn't happen. The events are totally
linked so they can't be independent.

This is fine for real numbers, that are commutative, but not with
clifford numbers. There is however, a extreme case where the
above theorem is true even in standard probability theory. Suppose
that $a,b$ and $c$ are three mutually exclusive propositions. Then,
$\psi(ab,c)=\psi(a,c)=\psi(b,c)=\psi(a,c)\psi(b,c)=0$ and we would
have to say that $a$ and $b$ are independent given $c$ even though
neither $a$ nor $b$ are possible given $c$. Anticommutativity allows
this to happen even when $\psi(a,c)$ and $\psi(b,c)$ are not zero. Two
events can be completely linked (i.e. mutually exclusive) and at the
same time be logically independent from each other! This is as weird
as entanglement in quantum mechanics.

\section{Flipping $n$ coins}

If \A is a $\sigma$-algebra of propositions in $X$ then, by
the $\sigma$-additivity property, every
$\psi\in H(\A)$ is completely specified on $\A$ by just giving
$\psi(x)$ for all $x\in X$, i.e.,
\begin{equation}
\psi(x) = \sum_{y\in X} \psi(y)\delta_{y}(x) \label{eq:elpsi}
\end{equation}
where for $x,y\in X$, $\delta_{y}(x)$ is $1\in%
\G$ if $x=y$ and $0\in \G$ otherwise.

We consider the following special case.

\subsection{The Binomial experiment with $\psi$s}

Let $a$ be an arbitrary proposition and let $X=\{a,\bar{a}\}$ and $%
\A = \{1,0,a,\bar{a}\}$. Clearly $\A$ is a
boolean algebra of propositions in $X$. From (\ref{eq:elpsi}) we
have
\begin{equation}
\psi(x) = A\delta_{a}(x) + B\delta_{\bar{a}}(x)  \label{eq:elpsi1}
\end{equation}
where $\psi\in H(\A)$ and $A,B\in\G$. 
This is the canonical Bernoulli experiment. There are only two possible
outcomes $a$ and $\bar{a}$ with partial truths encoded by the clifford
numbers $A=\psi(a)$ and $B=\psi(\bar{a})$.
As in standard probability
theory, consider now $n$ independent repetitions of the Bernoulli
experiment. i.e., consider $X^{n}$ with its
corresponding boolean algebra $\A^{n}$ of elements in $X^{n}$
(see (\ref{eq:An})). 
From (\ref{eq:elpsi}) a general $\psi_{n}\in H(\A^{n})$ is given by, 
\begin{equation}
\psi_{n}(x) = \sum_{y\in X^{n}}\psi_{n}(y)\delta_{y}(x)  \label{eq:psin}
\end{equation}
From the assumption that $\psi_{n}$ make the different repetitions independent,
we obtain, using (\ref{eq:prodrule}) that
\begin{equation}
\psi_{n}(x) = \psi_{n}(x_{1},\dots,x_{n}) = M_{n}(m(x))  \label{eq:Mn}
\end{equation}
where for each integer $k$ with $0\le k\le n$, $M_{n}(k)\in\G$
is the symmetrization of the product $A^{k}B^{n-k}$ and $m(x)$ is the number
of $a$'s in $x$. Here are some examples for $n=2$ and $n=3$, 
\begin{eqnarray*}
\psi_{2}(a,a) &=& A^{2} = M_{2}(2),\ \ \psi_{2}(a,\bar{a})= \psi_{2}(\bar{a}%
,a) = \frac{1}{2}[AB + BA] = M_{2}(1) \\
\psi_{3}(a,a,\bar{a}) &=& \psi_{3}(a,\bar{a},a)=\psi_{3}(\bar{a},a,a) = 
\frac{1}{3}[A^{2}B + ABA + BA^{2}] = M_{3}(2).
\end{eqnarray*}

Now define the proposition $P_{n}^{k}\in \A^{n}$ by, 
\begin{equation}
P_{n}^{k} = \mbox{``exactly $k$ of the $n$ repetitions is an $a$''}
\label{eq:Pnk}
\end{equation}
Recall that by (\ref{eq:bhat}) we have, 
\begin{equation}
P_{n}^{k}\psi_{n}(x) = \psi_{n}(P_{n}^{k}x) = \left\{ 
\begin{array}{ll}
\psi_{n}(x) & \mbox{if \ } P_{n}^{k}x \neq 0 \\ 
0 & \mbox{otherwise}
\end{array}
\right.
\end{equation}
By the first part of theorem (1) we have that $\psi_{n}$ makes $P_{n}^{k}$
true when $P_{n}^{k}\psi_{n}=\psi_{n}$. So the question is: \emph{How far is 
$\psi_{n}$ from making $P_{n}^{k}$ true?}. Answer: 
$\|\psi_{n} - P_{n}^{k}\psi_{n} \|^{2}$.

\subsection{Computation of $\|\psi_{n} - P_{n}^{k}\psi_{n} \|^{2}$}
To compute this distance we use the fact that $P_{n}^{k}$, and its
negation in $\A^{n}$, $1-P_{n}^{k}$, are mutually exclusive
propositions hence orthogonal (see (\ref{eq:abpythagoras})) and by
pythagoras,
\begin{equation}
\Vert \psi _{n}-P_{n}^{k}\psi _{n}\Vert ^{2}=\Vert \psi _{n}\Vert ^{2}-\Vert
P_{n}^{k}\psi _{n}\Vert ^{2}  \label{eq:pythagoras}
\end{equation}
Let us compute each of these terms. From (\ref{eq:innerprod}), 
\begin{equation}
\Vert \psi _{n}\Vert ^{2}=\sum_{x\in X^{n}}\langle \psi _{n}^{\dagger
}(x)\psi _{n}(x)\rangle _{0}.  \label{eq:norm}
\end{equation}
and using (\ref{eq:psin}) and (\ref{eq:Mn}) we can write, 
\begin{equation}
\psi _{n}(x)=\sum_{y\in X^{n}}M_{n}(m(y))\delta _{y}(x)  \label{eq:sumMn}
\end{equation}
from where we obtain, 
\begin{eqnarray*}
\psi _{n}^{\dagger }(x)\psi _{n}(x) &=&\sum_{y_{1},y_{2}\in
X^{n}}M_{n}^{\dagger }(m(y_{1})M_{n}(m(y_{2}))\delta _{y_{1}}(x)\delta
_{y_{2}}(x) \\
&=&\sum_{x\in X^{n}}M_{n}^{\dagger }(m(x))M_{n}(m(x)) \\
&&
\end{eqnarray*}
and replacing in (\ref{eq:norm}) we get, 
\begin{eqnarray}
\Vert \psi _{n}\Vert ^{2} &=&\sum_{x\in X^{n}}\left| M_{n}(m(x))\right| ^{2}
\nonumber \\
\Vert \psi _{n}\Vert ^{2} &=&\sum_{j=0}^{n}{{n}\choose{j}}|M_{n}(j)|^{2}
\label{eq:psin2}
\end{eqnarray}
the last equation followed from the fact that there are ${{n}\choose{j}}$
propositions in $X^{n}$ with exactly $j$ components equal to $a$. We use the
same fact again to compute the other norm in (\ref{eq:pythagoras}), 
\begin{equation}
\Vert P_{n}^{k}\psi _{n}\Vert ^{2}=\sum_{x\in X^{n}}\langle \psi
_{n}^{\dagger }(P_{n}^{k}x)\psi _{n}(P_{n}^{k}x)\rangle _{0}
\end{equation}
to obtain, 
\begin{equation}
\Vert P_{n}^{k}\psi _{n}\Vert ^{2}={{n}\choose{k}}|M_{n}(k)|^{2}
\label{eq:npn2}
\end{equation}
Replacing (\ref{eq:psin2}) and (\ref{eq:npn2}) in (\ref{eq:pythagoras}) we
get, 
\begin{equation}
\Vert \psi _{n}-P_{n}^{k}\psi _{n}\Vert ^{2}=\sum_{j=0}^{n}{{n}\choose{j}}%
|M_{n}(j)|^{2}-{{n}\choose{k}}|M_{n}(k)|^{2}  \label{eq:pythagoras2}
\end{equation}
Let us consider the proposition, $P_{n,\epsilon }^{f}\in \A%
^{n}$ defined by, 
\begin{eqnarray}
P_{n,\epsilon }^{f} &=&%
\mbox{``The observed frequency of $a$'s in $n$
 independent repetitions is}  \nonumber \\
&&\mbox{$k/n$ with $f-\epsilon\le \frac{k}{n}\le f+\epsilon$''}
\end{eqnarray}
in other words for $x\in X^{n}$, $P_{n,\epsilon }^{f}x\neq 0$ when and only
when the proportion of $a$'s in $x=(x_{1},\ldots ,x_{n})$ is within $%
\epsilon $ from the specified frequency $f$. The proposition $P_{n,\epsilon
}^{f}$ is equal to the following disjunction of $2n\epsilon +1$ mutually
exclusive propositions $P_{n}^{k}$: 
\begin{equation}
P_{n,\epsilon }^{f}=\sum_{k=n(f-\epsilon )}^{n(f+\epsilon )}P_{n}^{k}
\label{eq:pnef}
\end{equation}
hence, from (\ref{eq:abpythagoras}) we get, 
\begin{equation}
\Vert P_{n,\epsilon }^{f}\psi _{n}\Vert ^{2}=\sum_{k=n(f-\epsilon
)}^{n(f+\epsilon )}\Vert P_{n}^{k}\psi _{n}\Vert ^{2}
\label{eq:pfpsi2}
\end{equation}
and from (\ref{eq:pythagoras}) and (\ref{eq:pythagoras2}) we can write, 
\begin{equation}
\Vert \psi _{n}-P_{n,\epsilon }^{f}\psi _{n}\Vert ^{2}=\sum_{j=0}^{n}{
{n}\choose{j}}|M_{n}(j)|^{2}-\sum_{k=n(f-\epsilon )}^{n(f+\epsilon )}{{n}%
\choose{k}}|M_{n}(k)|^{2}  \label{eq:pythagoras3}
\end{equation}
In general this distance increases without limit as $n\rightarrow \infty $
but it can converge relative to the size of $\psi _{n}$. Let us define the
relative error by, 
\begin{equation}
\Delta _{n,\epsilon }^{f}=\frac{\Vert \psi _{n}-P_{n,\epsilon }^{f}\psi
_{n}\Vert ^{2}}{\Vert \psi _{n}\Vert ^{2}}  \label{eq:deltanef}
\end{equation}
Using (\ref{eq:pythagoras3}) and (\ref{eq:psin2}) we have,
\begin{equation}
\Delta _{n,\epsilon }^{f}= 1 - 
\frac{{\displaystyle 
 \sum_{k=n(f-\epsilon )}^{n(f+\epsilon )}{{n}\choose{k}}|M_{n}(k)|^{2}}}
     {{\displaystyle \sum_{k=0}^{n}{{n}\choose{k}}|M_{n}(k)|^{2}}}
		\label{eq:deltanef1}
\end{equation}
We separate the computation of $\Delta _{n,\epsilon }^{f}$ into three
different cases.

\subsection{Case: $AB=BA$}
From (\ref{eq:deltanef1}) we can write the following,
\begin{mytheorem}
If $AB\ne 0$, $AB=BA$ and $|A^{k}B^{n-k}| = |A|^{k}|B|^{n-k}$ then,
\begin{equation}
\Delta _{n,\epsilon }^{f}= 1 - 
\sum_{k=n(f-\epsilon )}^{n(f+\epsilon )}{{n}\choose{k}}p^{k}(1-p)^{n-k}
	\label{eq:deltapq}
\end{equation}
where,
\begin{equation}
p = \frac{|A|^{2}}{|A|^{2} + |B|^{2}} \label{eq:p}
\end{equation}
\end{mytheorem}
{\bf Proof:}\\
Under the conditions of the theorem we have,
\[ |M_{n}(k)|^{2} = |A|^{2k}|B|^{2(n-k)} \]
replacing this last equation in (\ref{eq:deltanef1}) and noticing that,
\[ \sum_{k=0}^{n}{{n}\choose{k}}|A|^{2k}|B|^{2(n-k)} = 
      \left( |A|^{2} + |B|^{2} \right)^{n} \]
we immediately obtain (\ref{eq:deltapq}) and (\ref{eq:p}) $\bullet$

It is not always true that for $A,B\in\G, |AB|=|A| |B|$ even when
$AB=BA$ (take for example $A = 1+\alpha u, B=1-\beta u$ for a unit
vector $u$ and scalars $\alpha$ and $\beta$) so the extra condition
besides commutativity is needed for the theorem to be true.

\subsection{Case: $AB=0$}
Unlike the real (or complex) numbers, the product of non zero clifford
numbers can be zero (e.g. take $\alpha = \beta = 1$ in the example
above) so this case is not trivial.  When $AB=0$ the following is true,
\begin{mytheorem}
If $AB=0$ then,
\begin{equation}
\Delta _{n,\epsilon }^{f} = \left\{ \begin{array}{ll}
	\frac{{\displaystyle |A|^{2n}}}{{\displaystyle |A|^{2n}+|B|^{2n}}} 
	& \mbox{if $f=0$} \vspace{10pt}\\
	1 & \mbox{if $0<f<1$} \vspace{10pt}\\
	\frac{{\displaystyle |B|^{2n}}}{{\displaystyle |A|^{2n}+|B|^{2n}}} 
	& \mbox{if $f=1$}
			\end{array}
	\right.  \label{eq:deltaba0}
\end{equation}
\end{mytheorem}
{\bf Proof:}\\
Notice that when $AB=0$ then all the symmetrized products, except the
two extremes are zero, i.e., $M_{n}(k)=0$ for all $0<k<n$ and
$M_{n}(n)=A^{n}$ and $M_{n}(0)=B^{n}$. Substituting these values into
(\ref{eq:deltanef1}) we obtain (\ref{eq:deltaba0}) $\bullet$

\subsection{Case: $AB = -BA$}
When $A$ and $B$ anticommute we have,

\begin{mytheorem}
If $AB\ne 0$, $AB = -BA$ and $|A^{k}B^{n-k}| = |A|^{k}|B|^{n-k}$ then,
\begin{equation}
\Delta _{n,\epsilon }^{f} = 1 - \frac{{\displaystyle 
 \sum_{k=n(f-\epsilon)}^{n(f+\epsilon)}b_{n}(k)(1-2\lambda_{n}(k))^{2}}}
{{\displaystyle
 \sum_{k=0}^{n}b_{n}(k)(1-2\lambda_{n}(k))^{2}}} \label{eq:antiab}
\end{equation}
where $b_{n}(k)$ are the binomial probabilities,
\begin{equation}
b_{n}(k) = {{n}\choose{k}}p^{k}(1-p)^{n-k}, \mbox{\ with $p$ as before. i.e.,\ }
p = \frac{|A|^{2}}{|A|^{2}+|B|^{2}}  \label{eq:bnk}
\end{equation}
and the numbers $\lambda_{n}(k)$ satisfy $\lambda_{n}(k) = \lambda_{n}(n-k)$ and
for $k\le n/2$, $\lambda_{n}(k)$ is the chance of drawing and odd number of
RED balls out of $k$ draws without replacement from a box containing either:
$n/2$ REDS and $n/2$ BLUES if $n$ is even or $(n+1)/2$ REDS and $(n-1)/2$ BLUES
if $n$ is odd.
\end{mytheorem}
{\bf Proof:}\\
Recall that $M_{n}(k)$ is the symmetrization of $A^{k}B^{n-k}$, i.e., the
average over all the permutations of $A^{k}B^{n-k}$. There are ${{n}\choose{k}}$
permutations and, by the assumed anticommutativity of $A$ with $B$, each 
permutation is either $A^{k}B^{n-k}$ or $-A^{k}B^{n-k}$ so we have,
\begin{equation}
M_{n}(k) = \frac{\rho(n,k)A^{k}B^{n-k}}{{{n}\choose{k}}} \label{eq:rho}
\end{equation}
where $\rho(n,k)$ is an integer. From the fact that $|M_{n}(k)|$ is
invariant under the transformation: $A\rightarrow B$, $B\rightarrow A$, and
$k\rightarrow (n-k)$ it follows that $|\rho(n,k)|=|\rho(n,n-k)|$.
In order to prove the theorem it is 
sufficient to show that,
\begin{equation}
\frac{|\rho(n,k)|}{{{n}\choose{k}}} = |1 - 2\lambda_{n}(k)|  \label{eq:rhonk}
\end{equation}
since if (\ref{eq:rhonk}) is true, by using the conditions of the theorem we have,
\begin{equation}
|M_{n}(k)|^{2} = (1-2\lambda_{n}(k))^{2} |A|^{2k}|B|^{2(n-k)} \label{eq:rhonk1}
\end{equation}
and dividing the numerator and the denominator of (\ref{eq:deltanef}) by
$(|A|^{2}+|B|^{2})^{n}$ we obtain (\ref{eq:antiab}).

Let us show that (\ref{eq:rhonk}) is true by giving an explicit
formula for $|\rho(n,k)|$ when $k\le n/2$. To do this, represent each
permutation of $A^{k}B^{n-k}$ by the $k$ integers $(j_{1}j_{2}\ldots
j_{k})$ that correspond to the positions of the $A$'s in increasing
order. For example, for $n=6$ and $k=3$, the permutation $ABABBA$ is
represented by $(136)$, since the $A$s are found at positions $1,3$
and $6$. The permutation $AABBBA$ is represented by $(126)$ etc. Define
the parity of $(j_{1}\ldots j_{k})$ as
\begin{equation}
\mbox{parity of \ } (j_{1}j_{2}\ldots j_{k}) = (-1)^{j_{1}+j_{2}\ldots +j_{k}}
= (-1)^{j_{1}}(-1)^{j_{2}}\ldots (-1)^{j_{k}}
  \label{eq:parity}
\end{equation}
Note that the transposition of an $A$ with a $B$, located next to it,
changes by one the position of that $A$ in the permutation and hence,
the parity of the permutation obtained after the transposition is
always the reverse of the parity of the original permutation.  From
this and the fact that we can transform any permutation into any other
by a sequence of transpositions it follows that two permutations have the
same parity if and only if the number of flips (transpositions)
necessary for transforming one permutation into the other is even.

The permutation $A^{k}B^{n-k}$ always corresponds to $(12\ldots k)$
and therefore an arbitrary permutation $(j_{1}j_{2}\ldots j_{k})$ will
have the same parity as $A^{k}B^{n-k}$ if the parity of the number of
odd integers in the set $\{j_{1},j_{2},\ldots,j_{k}\}$ is the same as
the parity of the number of odd integers in the set
$\{1,2,\ldots,k\}$. In other words, if there are an even number of odd
integers in the set $\{1,2,\ldots,k\}$ then every permutation
$(j_{1}j_{2}\ldots j_{k})$ which also contains an even number of odd
integers can be reorder into $A^{k}B^{n-k}$ but if the number of odd
integers in $\{j_{1},\ldots,j_{k}\}$ is odd then the permutation
reorders into $-A^{k}B^{n-k}$. Therefore, we can write

\begin{equation}
|\rho(n,k)| = |\sum_{1\le j_{1}< j_{2}\ldots < j_{k}\le n} 
(-1)^{j_{1}+j_{2}+\ldots + j_{k}}|
\label{eq:rhoequal}
\end{equation}
Thus, if we call $N_{e}$ the number of permutations with an even number
of odd integers among $\{j_{1}\ldots,j_{k}\}$ and we call $N_{o}$ the
number of permutations with an odd number of odds, then,
\begin{equation}
|\rho(n,k)| = | N_{e} - N_{o} | \label{eq:NeNo}
\end{equation}
using the fact that $N_{e}+N_{o} = {{n}\choose{k}}$ we also have that,
\begin{equation}
|\rho(n,k)| = |{{n}\choose{k}} - 2 N_{o}|  \label{eq:rho2}
\end{equation}
We now turn to the computation of $N_{o}$. Let $N_{o}(m)$ be the total 
number of permutations $(j_{1}j_{2}\ldots j_{k})$ with exactly $m$
of the positions of the $A$'s being odd. We have,
\begin{equation}
N_{o} = \left\{\begin{array}{ll}
{\displaystyle \sum_{t=0}^{\frac{k}{2}-1} N_{o}(2t+1)} & 
    	\mbox{ if $k$ is even} \vspace{10pt}\\
{\displaystyle \sum_{t=0}^{\frac{k-1}{2}} N_{o}(2t+1)} & 
	\mbox{ if $k$ is odd}
\end{array} \label{eq:No}
\right.
\end{equation}
where, for $0\le m\le k\le n/2$
\begin{equation}
N_{o}(m) = \left\{\begin{array}{ll}
 {{n/2}\choose{m}} {{n/2}\choose{k-m}} & \mbox{if $n$ is even} \vspace{10pt}\\
 {{(n+1)/2}\choose{m}} {{(n-1)/2}\choose{k-m}} & \mbox{if $n$ is odd}
\end{array} \label{eq:Nom}
\right.
\end{equation}
this is because the set $\{1,2,\ldots,n\}$ contains an equal number of odd
and even numbers when $n$ is even but the number of odds is one more than
the number of even when $n$ is odd.
So dividing (\ref{eq:rho2}) by ${{n}\choose{k}}$ and using (\ref{eq:No}) and
(\ref{eq:Nom}) we obtain (\ref{eq:rhonk}) with $\lambda_{n}(k)$ defined as
the theorem says. There are four different formulas for $\lambda_{n}(k)$
depending on the parities of $n$ and $k$.Let us check one of them.
When $n$ and $k$ are both even and $k\le n/2$ we have,
\begin{equation}
\lambda_{n}(k) = \sum_{t=0}^{\frac{k}{2}-1}\frac{{\displaystyle
	 {{n/2}\choose{2t+1}} {{n/2}\choose{k-2t-1}} }}
	{{\displaystyle {{n}\choose{k}} }}	\label{eq:lambda}
\end{equation}
and we can see that (\ref{eq:lambda}) is the chance of drawing an odd
number of red balls when drawing at random $k$ balls, without
replacement, from a box containing $n/2$ red balls and $n/2$ blue
balls.  This completes the proof of the theorem $\bullet$

\section{The weak law of large numbers}

\subsection{Taking limits as $n\rightarrow\infty$}
In this section we compute 
\[\lim _{n\rightarrow \infty }\Delta_{n,\epsilon}^{f}\]
for the three cases considered in the previous section. 
\begin{mytheorem}
If $AB\neq 0$, $|A^{k}B^{n-k}| = |A|^{k}|B|^{n-k}$ and either $AB=BA$
or $AB = -BA$ then, for all sufficiently small $\epsilon > 0$,
\begin{equation}
\lim_{n\rightarrow\infty}\Delta_{n,\epsilon}^{f} = \left\{
	\begin{array}{ll}
	0 & \mbox{if $f = p$} \\
	1 & \mbox{if $f \neq p$}
	\end{array}
\right. \label{eq:limdelta}
\end{equation}
where as before, $p =\frac{|A|^{2}}{|A|^{2}+|B|^{2}}$.
Moreover, if $AB = 0$, then $\forall \epsilon > 0$,
\begin{equation}
\lim_{n\rightarrow\infty}\Delta_{n,\epsilon}^{f} = \left\{
	\begin{array}{ll}
	0 & \mbox{if ($f=0$ and $|A| < |B|$) or ($f=1$ and $|A|>|B|$)} 
 		\vspace{10pt}\\
	1/2 & \mbox{if $|A| = |B|$ and either $f=0$ or $f=1$}
		\vspace{10pt}\\
	1 & \mbox{otherwise}
	\end{array}
\right. \label{eq:limdelta1}
\end{equation}
\end{mytheorem}
{\bf Proof}\\ 
For the first part we use equations, (\ref{eq:deltapq})
and (\ref{eq:antiab}). By the usual gaussian approximation for the
binomial probabilities (e.g. see \cite{shiryayev84} p.59) we have 
that for any integers 
$0\le k_{1}\le k_{2}\le n$ and any function $g_{n}$ with finite expectation
with respect to the standard gaussian,
\begin{equation}
\sum_{k=k_{1}}^{k_{2}} b_{n}(k) g_{n}(k) =
 \int_{\frac{k1-np}{\sqrt{npq}}}^{\frac{k2-np}{\sqrt{npq}}}
  g_{n}(np + x \sqrt{npq}) \frac{1}{\sqrt{2\pi}}e^{\frac{-x^2}{2}} dx
   \left(1+o(n^{0})\right)   \label{eq:gaussapprox}
\end{equation}
thus, taking $k_{1} = n(f-\epsilon)$, $k_{2} = n(f+\epsilon)$,
$g_{n}(y)= 1$ for $0\le y \le n$ and $g_{n}(y)=0$ outside $[0,n]$
we obtain from equation (\ref{eq:deltapq}) that,
\begin{equation}
\lim_{n\rightarrow\infty}\Delta_{n,\epsilon}^{f} = 
 1 - \lim_{n\rightarrow\infty}
\int_{\frac{n(f-p-\epsilon)}{\sqrt{npq}}}^{\frac{n(f-p+\epsilon)}{\sqrt{npq}}}
 \frac{1}{\sqrt{2\pi}}e^{\frac{-x^2}{2}} dx   \label{eq:gaussapprox1}
\end{equation}
hence, when $f\ne p$ for any $0 < \epsilon < |f-p|$ the limits of the
integral in equation (\ref{eq:gaussapprox1}) are both positive or both
negative and both going to $\infty$ as $n\rightarrow\infty$ so the desired
limit is $1-0=0$. On the other hand when $f=p$ for any $\epsilon > 0$
the desired limit is $1-1 =0$ and this shows that
(\ref{eq:limdelta}) is true for the commutative case. 
To show (\ref{eq:limdelta}) for the anticommutative case we take
\begin{equation}
g_{n}(y) = \frac{1}{4}y^{2}(y-1)^{2}
	\label{eq:gnanti}
\end{equation}
which increases like $y^{6}$ and therefore it has finite expectation
with respect to the standard gaussian. If we show that
\begin{equation}
g_{n}(k) = n^{2}(1-2\lambda_{n}(k))^{2} + o(n^{0})
               \label{eq:gnanti1}
\end{equation}
then it will follow from (\ref{eq:gnanti1}), (\ref{eq:gaussapprox}) and
(\ref{eq:antiab}) that,
\begin{equation}
\lim_{n\rightarrow\infty}\Delta_{n,\epsilon}^{f} =
1 - \lim_{n\rightarrow\infty}
\frac{ 
 \int_{\frac{n(f-p-\epsilon)}{\sqrt{npq}}}^{\frac{n(f-p+\epsilon)}{\sqrt{npq}}}
 g_{n}(np + x \sqrt{npq}) \frac{1}{\sqrt{2\pi}}e^{\frac{-x^2}{2}} dx}
{\int_{\frac{-np}{\sqrt{npq}}}^{\frac{nq}{\sqrt{npq}}}
 g_{n}(np + x \sqrt{npq}) \frac{1}{\sqrt{2\pi}}e^{\frac{-x^2}{2}} dx}
   \label{eq:gnanti2}
\end{equation}
and by the same reasoning as in the commutative case we
obtain (\ref{eq:limdelta}) for the anticommutative case.
Let us then show (\ref{eq:gnanti1}). Notice that from (\ref{eq:lambda})
we can write,
\begin{equation}
\lambda_{n}(k) = \sum_{t=0}^{\frac{k}{2}-1}W(n,2t+1,k) \label{eq:lambdaw}
\end{equation}
where the hypergeometric probabilities,
\begin{eqnarray}
W(n,m,k) & = & \frac{{\displaystyle
	 {{n/2}\choose{m}} {{n/2}\choose{k-m}} }}
	{{\displaystyle {{n}\choose{k}} }}  \nonumber \\
 & = & \frac{{\displaystyle
	\left[\frac{1}{n^{m}}{{n/2}\choose{m}}\right]
        \left[\frac{1}{n^{k-m}}{{n/2}\choose{k-m}}\right] }}
	{{\displaystyle \left[\frac{1}{n^{k}}{{n}\choose{k}}\right] }} \\ 
 & = & \frac{\left[
 \frac{1}{m!}\frac{1}{2}(\frac{1}{2}-\frac{1}{n})
\cdots (\frac{1}{2}-\frac{m-1}{n})  \right]
\left[
 \frac{1}{(k-m)!}\frac{1}{2}(\frac{1}{2}-\frac{1}{n})
\cdots (\frac{1}{2}-\frac{k-m-1}{n})  \right]}
{\left[
 \frac{1}{k!}1(1-\frac{1}{n})(1-\frac{2}{n})
\cdots (1-\frac{k-1}{n})  \right]} \nonumber
\end{eqnarray}
expanding the products up to terms of order 
$(1/n)$ and letting $W=W(n,m,k)$ we have,
\begin{eqnarray}
W & = & {{k}\choose{m}} \frac
{\left[ 2^{-m}\{1 - \frac{m(m-1)}{n} + o(n^{-1})\} \right]
 \left[ 2^{m-k}\{1-\frac{(k-m)(k-m-1)}{n} + o(n^{-1})\} \right]}
{1 - \frac{k(k-1)}{n} + o(n^{-1})} \nonumber \\
 & = & {{k}\choose{m}} \left(\frac{1}{2}\right)^{k}
\left\{ 1 - \frac{1}{n}[m^{2}+(k-m)^{2}-k] + o(n^{-1})\right\}
\left\{ 1 + \frac{k(k-1)}{n} + o(n^{-1}) \right\} \nonumber \\
 & = & {{k}\choose{m}} \left(\frac{1}{2}\right)^{k}
\left\{ 1 + \frac{2}{n}m(k-m) + o(n^{-1})\right\} \label{eq:Wexp}
\end{eqnarray}
We can readily check that,
\begin{equation}
\sum_{t=0}^{\frac{k}{2}-1} {{k}\choose{2t+1}}
    \left(\frac{1}{2}\right)^{k} = \frac{1}{2} \label{eq:sum12}
\end{equation}
and that,
\begin{equation}
\sum_{t=0}^{\frac{k}{2}-1}(2t+1)(k-2t-1){{k}\choose{2t+1}}
 = \frac{1}{4}k(k-1)2^{k-1} \label{eq:sum2}
\end{equation}
From (\ref{eq:sum12}), (\ref{eq:sum2}), (\ref{eq:Wexp}) and 
(\ref{eq:lambdaw}) we have,
\begin{equation}
\lambda_{n}(k) - \frac{1}{2} = \frac{k(k-1)}{4n} + o(n^{-1}). 
	\label{eq:lnk12}
\end{equation}
Squaring both sides of (\ref{eq:lnk12}) and multiplying
through by $4n^{2}$ we obtain,
\begin{equation}
n^{2}(1-2\lambda_{n}(k))^{2} = \frac{1}{4}k^{2}(k-1)^{2} + o(n^{0})
	\label{eq:n212}
\end{equation}
which is exactly (\ref{eq:gnanti1}). This ends the proof
for the anticommutative case. The second part of the theorem
i.e. (\ref{eq:limdelta1}) follows directly from (\ref{eq:deltaba0})
by taking limits as $n\rightarrow\infty \bullet$

\subsection{Flipping an infinite number of coins}

As in standard probability theory there is a subtle nuisance with
limits such as (\ref{eq:limdelta}) and (\ref{eq:limdelta1}) that needs
to be faced in order to have a straight probabilistic interpretation
for laws of large numbers. The problem with (\ref{eq:limdelta}) and
(\ref{eq:limdelta1}) is that it is not clear how to paste all the
$\psi_{n}$ together into one global $\psi_{\infty}$. It was 
due to these kind of problems that modern measure-theoretic
probability theory was born.

To be able to make statements about infinite sequences of bernoulli
trials we need to specify a boolean $\sigma$-algebra, $\A^{\infty}$,
that contains at least those statements. This can be done as in
standard probability theory (e.g. see \cite{chow88}), i.e.
$\A^{\infty}$ is defined as the smallest $\sigma$-algebra containing
the cylinder sets, in particular it contains the propositions
$P_{n}^{k}$ defined in (\ref{eq:Pnk}) but now $n$ refers to the first
$n$ repetitions in an infinite sequence of bernoulli trials.  Having
constructed $\A^{\infty}$ we also need to construct the Hilbert space,
$H(\A^{\infty})$, containing the functions $\psi=\psi_{\infty}$.
Again, the construction is not trivial but well known in functional
analysis as the standard construction of an infinite tensor product of
Hilbert spaces (e.g. see \cite{hartle68}).  These standard
constructions allow us to write,
\begin{equation}
P_{n}^{k}\psi = P_{n}^{k}\psi_{n} \label{eq:Pinfty}
\end{equation}
where $\psi = \psi_{\infty}\in H(\A^{\infty})$.  Equation
(\ref{eq:Pinfty}) can be used to re-write the statements 
(\ref{eq:limdelta}) and (\ref{eq:limdelta1}) as,
\begin{mytheorem}
Let $X^{\infty}$ be the space of infinite sequences of independent
tosses of a coin and let $\A^{\infty}$ be the smallest
$\sigma$-algebra containing all the propositions $P_{n}^{k}$ about
elements in $X^{\infty}$.  If for each toss the $\psi$ values for
falling heads and tails are the clifford numbers $A$ and $B$
satisfying,
\begin{enumerate}
\item $|A|^{2}+|B|^{2} = 1$
\item $AB\ne 0$
\item either $AB = BA$ or $AB = -AB$
\item $|A^{k}B^{n-k}| = |A|^{k}|B|^{n-k}\ \ 
         \forall n\in \N, \forall 0 \le k \le n$.
\end{enumerate}
Then, for all sufficiently small $\epsilon > 0$ the propositions,
\[P_{\infty,\epsilon}^{|A|^{2}}\in \A^{\infty}\] 
are true.
\end{mytheorem}
{\bf Proof}\\
Under the conditions of the theorem we have from (\ref{eq:Pinfty}) and
(\ref{eq:limdelta}) that when the $\psi_{n}$ are normalized i.e. when
$\|\psi_{n}\|=1$ for all $n$ then,
\[ \|\psi - P_{n,\epsilon}^{|A|^{2}}\psi \| \rightarrow 0 \ \mbox{as\  } 
   n \rightarrow\infty \]
or equivalently,
\begin{equation}
\lim_{n\rightarrow\infty} P_{n,\epsilon}^{|A|^{2}} \psi =
 P_{\infty,\epsilon}^{|A|^{2}}\psi = \psi	\label{eq:limnr}
\end{equation}
so that $\psi$ is an eigen vector of the operator 
$P_{\infty,\epsilon}^{|A|^{2}}$ with eigen value $1$ and thus, it makes
the proposition true $\bullet$

We also have,
\begin{mytheorem}
Let $X^{\infty}$ and $\A^{\infty}$ be as in the previous theorem but
now suppose that the clifford numbers $A$ and $B$ satisfy,
\begin{enumerate}
\item $AB = 0$
\item $|A| > |B|$
\end{enumerate}
Then for all $\epsilon > 0$ the propositions,
\[P_{\infty,\epsilon}^{1}\in \A^{\infty}\] 
are true.
\end{mytheorem}
{\bf Proof}\\
Under the conditions of the theorem we have from (\ref{eq:Pinfty}) and
(\ref{eq:limdelta1}) that when the $\psi_{n}$ are all of unit norm then
\[ \|\psi - P_{n,\epsilon}^{1}\psi \| \rightarrow 0 \ \mbox{as\  } 
   n \rightarrow\infty \]
or equivalently,
\begin{equation}
\lim_{n\rightarrow\infty} P_{n,\epsilon}^{1} \psi =
 P_{\infty,\epsilon}^{1}\psi = \psi	\label{eq:limnr1}
\end{equation}
so that $\psi$ is an eigen vector of the operator 
$P_{\infty,\epsilon}^{1}$ with eigen value $1$ and thus, it makes
the proposition true $\bullet$

\subsection{Interpretation and examples}

The previous two theorems can be interpreted as in standard
probability theory. They say that an infinite sequence of independent
tosses of a coin with $\psi($ heads $) = A$ and $\psi($ tails $)= B$
will have for sure (relative to $\psi$) a frequency of heads within
$\epsilon$ from $|A|^{2}$ in the first case and within $\epsilon$ from
$1$ in the $AB=0$ case.  When $AB=0$ the theorem assures us that
(again relative to $\psi$) the coin will show up heads with frequency
$100$\% whenever $|A| > |B|$ !  

The four conditions on $A$ and $B$ that are needed for the $AB\ne 0$
case, impose heavy restrictions on the possible values that $A$ and
$B$ can take but there are lots of examples. Let $p$ be a real number
in the interval $[0,1]$ and consider,
\begin{description}
\item[Example 1] 
	\begin{equation}
	A = \sqrt{p}\hspace{25pt} B = \sqrt{1-p}	\label{eq:E1}
	\end{equation}
\item[Example 2]
	\begin{equation}
	A = \sqrt{p}  \hspace{25pt} B = \sqrt{1-p}\hat{B} \label{eq:E2}
	\end{equation}
where $\hat{B} = \sigma_{1}\sigma_{2}\ldots\sigma_{r}$ is a unit blade, 
i.e. it can be factorized into a product of orthogonal (anticommuting)
unit vectors $\sigma_{j}$.
\item[Example 3]
	\begin{equation}
	A = \sqrt{p}\hat{A} \hspace{25pt} B = \sqrt{1-p}\hat{B} \label{eq:E3}
	\end{equation}
where $\hat{A}$ and $\hat{B}$ are both unit blades possibly of different
dimensions.
\item[Example 4]
	\begin{equation}
	A = \sqrt{p}\ e^{\I\alpha}\hat{A} \hspace{25pt} 
	B = \sqrt{1-p}\ e^{\I\beta}\hat{B} \label{eq:E4}
	\end{equation}
where $\alpha$ and $\beta$ are scalars,
$\hat{A}$ and $\hat{B}$ are both unit blades and $\I$ is any
multivector such that $\I^{2}=-1$ and $\I$ commutes or anticommutes
with both $\hat{A}$ and $\hat{B}$ i.e.  $\I\hat{A}=\pm\hat{A}\I$ and
$\I\hat{B}=\pm\hat{B}\I$

\end{description}
It can be readily check that all these examples satisfy the four
conditions of the theorem and hence, coin tosses with these $\psi$s will
show up heads with probability $p$.

\subsection{Why isn't every one a frequentist?}

For the same reason as in probability theory these laws of large
numbers can not be used to define what we mean by {\em the partial
truth that the coin will show up heads in the next toss} since the
theorem only says that the propositions $P_{\infty,\epsilon}^{p}$ are
made true {\em by } $\psi$. So any attempt to use the law of large
numbers as the definition of what $\psi$ is, or means, is therefore
circular.

\section{The Boolean algebra of Caticha's temporal filters}

Let $X$ be a set and let $\B$ be a $\sigma$-algebra of subsets of
$X$. Notice that we are using the standard set notation for the
elements of $\B$ instead of the logical notation used in the rest of
the paper. The reason for changing the notation is that the boolean
$\sigma$-algebra that we are trying to define is not $\B$ itself but
only based on $\B$. Think of $X$ as the set of possible locations
for a point particle and define the elementary propositions $e(x,t)$
by the statement: {\em the particle is at location $x$ at time $t$}.
As in \cite{caticha98}, $e(x,t)$ is a pure hypothesis not the
result of a measurement. The truth value of $e(x,t)$
can be obtained, at least in principle, by imagining a filter
that covers all of $X$ except at location $x$ where it has an
infinitesimal hole. This magical filter materializes only
for an instant at time $t$ and then disappears leaving no trace
of its existence. If after time $t$ we still find the particle
somewhere then we conclude that $e(x,t)$ is true. These filters
form a boolean algebra with the definitions below.

Let $T$ be a subset of the real line and define for
$t\in T$ and $B\in \B$ the proposition $e(B,t)$ as: {\em an elementary
filter at time $t$ with $B$ open}. Thus, $e(B,t)$ is true if and only
if the statement: {\em the particle is somewhere in $B$ at time $t$}
is true. We define the logical product of two elementary filters as
the operation of putting one on top of the other and we define the
negation of an elementary filter as the filter that closes the
holes and opens the rest. In symbols:
\begin{eqnarray}
e(B_{1},t)e(B_{2},t) & = & e(B_{1}\cap B_{2},t) \label{eq:etimese} \\
\overline{e(B,t)} & = & e(\overline{B},t) \label{eq:enot}\\
e(B_{1},t)+e(B_{2},t) & = & e(B_{1}\cup B_{2},t) \label{eq:epluse}
\end{eqnarray}
where $\overline{B} = X \setminus B$ is the complement of $B$ with respect to $X$.
Notice that (\ref{eq:epluse}) follows from (\ref{eq:etimese}) and 
(\ref{eq:enot}) by using De'Morgan's law i.e.,
\begin{eqnarray}
e(B_{1},t)+e(B_{2},t) & = & \overline{e(\overline{B_{1}},t)e(\overline{B_{2}},t)}
	\nonumber \\
 & = & \overline{e(\overline{B_{1}}\cap\overline{B_{2}},t)} \nonumber \\
 & = & e(B_{1}\cup B_{2},t)  \nonumber
\end{eqnarray}
We also have that for all $s,t\in T$,
\begin{eqnarray}
e(B_{1},s)e(B_{2},t) & = & \mbox{``Filter at time s followed 
             (or on top of) filter at time t''} \nonumber \\
e(B_{1},s)+e(B_{2},t) & = & \mbox{``Filter at 
           time s {\bf OR} filter at time t''} \nonumber \\
e(\phi,t) & = & \mbox{``Barrier (nothing open) at time $t$''} = 0 \\
e(X,t) & = & \mbox{``Absence of filter (all open) at time $t$''} = 1
\end{eqnarray}
We define $\F$ as the smallest $\sigma$-algebra containing the elementary
filters $e(B,t)$ i.e.,
\begin{equation}
\F = \sigma\left\{e(B,t) : B\in \B, t\in T \right\} \label{eq:Fdefined}
\end{equation}
The boolean algebra of temporal filters $\F$ is a spell out of the
usual algebra of events of a stochastic process with state space $X$.

\subsection{The Markov Property}

Due to the fact that there is no product rule for the unnormalized
 $\psi$s we cannot make use of the standard Markov property of
probability theory directly. The following definition is all that
is needed to recover non relativistic quantum mechanics,
\vspace{10pt}\\
{\bf Definition:}
$\psi\in H(\F)$ is said to have {\em independent segments given 
$c\in\F$} if for all
$n=1,2,\ldots$, all times $t_{1} < t_{2} < \ldots < t_{n}$
in $T$ and all locations $x_{1},x_{2},\ldots,x_{n}$ in $X$ the
propositions 
\[e(x_{1},t_{1})e(x_{2},t_{2}),\  e(x_{2},t_{2})e(x_{3},t_{3}),\ldots,
e(x_{n-1},t_{n-1})e(x_{n-1},t_{n-1})\] 
are independent given $c$.

\subsection{Time evolution and the Schr\"{o}dinger equation}

When $\psi\in H(\F)$ has independent segments, it evolves according
to the Schr\"{o}dringer equation. The usual jargon of quantum mechanics
is recovered with the notation,
\begin{description}
\item[Probability Amplitude:] $\psi(e(x,s)e(y,t),e(x_{0},t_{0}))$ is
the amplitude for the particle to go from location $x$ at time $s$ to
location $y$ at time $t > s$ given that it was initially prepared
at location $x_{0}$ at time $t_{0}$. We denote this amplitude
by $K(y,t;x,s)$.

\item[Wave Function:] $\psi(e(x,t),e(x_{0},t_{0}))$ is the amplitude of
going from the initial position to location $x$ at time $t$.  It is
often denoted by just $\Psi(x,t)$.
\end{description}
Thus, with this notation, a particle which is prepared by
$e(x_{0},t_{0})$ and for which $\psi\in H(\F)$ has independent
segments conditionally on this preparation, will satisfy,
\begin{equation}
\Psi(x,t) = \sum_{y\in X}\frac{1}{2}
 \left[ K(x,t;y,s)\Psi(y,s) + \Psi(y,s)K(x,t;y,s)\right] \label{eq:Schro}
\end{equation}
since
\begin{eqnarray*}
\Psi(x,t) & = & \sum_{y\in X}\psi(e(x,t)e(y,s),e(x_{0},t_{0})) \\
 & = & \sum_{y\in X}\psi
 \left( [e(x_{0},t_{0})e(y,s)]\ [e(y,s)e(x,t)],e(x_{0},t_{0})\right)
\end{eqnarray*}
taking derivatives in (\ref{eq:Schro}) with respect to $t$ and evaluating
at $t=s$ we obtain,
\[
  \left.\frac{\partial \Psi(x,t)}{\partial t}\right|_{t=s} 
              = \sum_{y\in X}\frac{1}{2}
 \left[ \left. \frac{\partial K(x,t;y,s)}{\partial t}\right|_{t=s}
\Psi(y,s) + \Psi(y,s)\left. 
 \frac{\partial K(x,t;y,s)}{\partial t}\right|_{t=s}\right]
\]
Defining the Hamiltonian $H$ by,
\begin{equation}
\left. \frac{\partial K(x,t;y,s)}{\partial t}\right|_{t=s} 
       = -\frac{i}{\hbar}H(x,y,s) \label{eq:hamiltonian}
\end{equation}
where $i$ is any multivector that squares to $-1$ and that it commutes
with all the $\psi$s. Relabeling $s$ with $t$ we can write
Schr{\"{o}}dinger equation for possible non-commuting $\psi$s as,
\begin{equation}
i\hbar\frac{\partial \Psi(x,t)}{\partial t}
              = \sum_{y\in X}\frac{1}{2}
 \left[ H(x,y,t)\Psi(y,t) + \Psi(y,t)H(y,x,t)\right]
  \label{eq:Schrodinger}
\end{equation}
when the wave functions $\Psi$ commute with the Hamiltonian, (e.g.
when all the $\psi$s take values in a commutative subspace of $\G$)
(\ref{eq:Schrodinger}) reduces to the usual Schr\"{o}dinger equation.

\section{Next:}
\begin{description}
\item[Using the Spacetime algebra] How to connect the above 
 with the Dirac-Hestenes equation.
\item[$\psi$ assignments in the real continuous case] Minimum
Fisher information and the Huber-Frieden derivation of the 
time independent Schr\"{o}dinger equation.
\item[$\psi$ and Brownian motion] Nagasawa's diffusion model.
\item[Comments and conclusion] What the hell is this all about
and  what it may be likely to become....
\end{description}

\bibliographystyle{plain}
\bibliography{QM}

\begin{thebibliography}{10}

\bibitem{aczel66}
J.~Aczel.
\newblock {\em Lectures on Functional equations and their appliations}.
\newblock Academic Press, New York., 1966.

\bibitem{caticha98}
A.~Caticha.
\newblock Consistency, amplitudes and probabilities in quantum theory.
\newblock {\em Phys. Rev.}, 1998.

\bibitem{chow88}
Y.~S. Chow and H.~Teicher.
\newblock {\em Probability Theory: Independence, Interchangeability,
  Martingales}.
\newblock Springer Texts in Statistics. Springer-Verlag, second edition, 1988.

\bibitem{cox46}
R.~T. Cox.
\newblock Probability, frequency and reasonable expectation.
\newblock {\em American Journal of Physics}, 14:1--13, 1946.

\bibitem{feynman48}
R.~P. Feynman.
\newblock {\em Rev. Mod. Phys.}, 20:267, 1948.

\bibitem{halmos74}
P.~R. Halmos.
\newblock {\em Measure Theory}, volume~18 of {\em GTM}.
\newblock Springer-Verlag, 1974.

\bibitem{hartle68}
J.~B. Hartle.
\newblock {\em Am. J. Phys.}, 36:704, 1968.

\bibitem{hestenes84}
D.~Hestenes and G.~Sobczyk.
\newblock {\em Clifford Algebra to Geometric Calculus}.
\newblock D. Reidel, 1984.

\bibitem{jaynesBook}
E.~T. Jaynes.
\newblock Probability theory: The logic of science.
\newblock http://omega.albany.edu:8008/JaynesBook.html.

\bibitem{kolmogorov33}
A.~N. Kolmogorov.
\newblock {\em Foundations of the Theory of Probability, 1933}.
\newblock Chelsea, New York, 1950.

\bibitem{smith90}
C.~R. Smith and G.~J. Erickson.
\newblock Probability theory and the associativity equation.
\newblock In P.~F. Fougere, editor, {\em Maximum Entropy and Bayesian Methods},
  pages 17--30. Dartmouth, USA, Kluwer, Academic publishers, 1990.

\bibitem{youssef91}
S.~Youssef.
\newblock {\em Mod. Phys. Lett.}, 6:225, 1991.

\end{thebibliography}

\end{document}